\documentclass[12pt]{article}
\usepackage{amsmath,amsthm,amsfonts}
\usepackage{url}
\newcommand{\pf}{\noindent {\bf Proof:} }
\newtheorem{theorem}{Theorem}
\newtheorem{remark}{Remark}

\newtheorem{corollary}[theorem]{Corollary}
\newtheorem{lemma}[theorem]{Lemma}

\title{Infinite ternary square-free words concatenated from permutations of a single word}
\author{James Currie\thanks{The author is 
supported by an NSERC Discovery Grant.}\\
Department of Mathematics and Statistics \\
University of Winnipeg \\
515 Portage Avenue \\
Winnipeg, Manitoba R3B 2E9 (Canada) \\
\url{j.currie@uwinnipeg.ca} }
\begin{document}
\date{\today}
\begin{abstract}
We answer a question of Harju: An infinite square-free ternary word with an $n$-stem factorization exists for any $n\ge 13$.

We show that there are uniform ternary morphisms of length $k$ for every $k\ge 23$. This resolves almost completely a problem of the author and Rampersad.
\end{abstract}
\maketitle
\section{Introduction}
Fixed points of morphisms have been used to create words avoiding powers and patterns since the  turn of the last century \cite{berstelthue}. Analysis of morphisms is simpler when the morphisms are highly symmetrical. Morphisms of Thue, Pleasants and Ker\"{a}nen \cite{berstelthue,pleasants,keranen} , for example, may be expressed on alphabets $\{0,1,\ldots, n\}$ such that the image of each letter $i$ is obtained from the image of 0 using the cyclic shift $0\rightarrow 1\rightarrow 2\cdots\rightarrow n\rightarrow 0$, repeated $i$ times. Evidently, these morphisms are $k$-uniform for some $k$, and their fixed points are obtained by concatenating blocks of length $k$, each of which is equivalent to the first block under some permutation of alphabet letters.

Following Harju \cite{harju}, we say that a finite word $p$ over alphabet $\Sigma$  is a {\bf stem} of length $n$ of $w$ if 
$$w =  pw_1w_2w_3\cdots $$
where for each $i$
there exists a morphism $\mu_i$ on $\Sigma$ extending a permutation of the letters of $\Sigma$, with $w_i = \mu_i(p)$. In this case the word $w$ has an {\bf $n$-stem factorization}.

Harju poses this problem: Are there infinite square-free ternary words with an $n$-stem factorization for each $n\ge 13$?

In the paper of Harju it is already shown that such words exist for $n=14, 15, 16$. Mike M\"{u}ller has pointed out to the author \cite{muller} the existence of such words for $n =20, 21$ and $22$:

For $n=20$, a square-free word with a stem of length 20 is generated by the morphism

\begin{eqnarray*}
h(0)&=&012102010210121021202101202120121012010212021012102120210201\\
&&\mbox{-}21012021201210120102\\
h(1)&=&012102010210121021202101202120121012010212021012102120210201\\&&\mbox{-}021201020120212012102010212021020102101210201210120102012021\\
h(2)&=&012102010210121021201020121012010201202120102120210201021012\\&&\mbox{-}10201210120102012021
\end{eqnarray*}

It is squarefree by Crochemore's characterization and each image is a concatenation of images of the stem 01210201021012102120 under various permutations.

For $n=21$, one could take

\begin{eqnarray*}
        h(0)&=&012021020102120102012102120121012021012102120102101210201210
\\&&\mbox{-}1202102012021201021202102012102120210120212012102012021201021
\\&&\mbox{-}20210120102101210201210120102120121012021012102\\
        h(1)&=&012021020102120102012021012010201210201021201210212021012021
\\&&\mbox{-}201210201202120102120210120102101210201210120210201202120102
\\&&\mbox{-}120210201210212021012021201021012010201210201021\\
       h(2)&=&012021020102120102012021012010201210201021201210212021012021
\\&&\mbox{-}201021012010201210201021012021020102120102012102120121012021
\\&&\mbox{-}012102120102101210201210120102120121012021012102
\end{eqnarray*}

For $n=22$, here is a solution from M\"{u}ller:

\begin{eqnarray*}
        h(0)&=&012021020102120210201210212012101202120121021201021012102
\\&&\mbox{-}010210120210201202120102012021020121021202101210212010210
\\&&\mbox{-}120102012101201021\\
h(1)&=&0120210201021202102012102120121012021201210212010210121020102
\\&&\mbox{-}1012021020120212010201202101201021012102010210120102120121
\\&&\mbox{-}0120212012102\\
h(2)&=&01202102010212021020121021201210120212012102120102101210201021
\\&&\mbox{-}0120102120121012021201210201202102010212021020120210120102
\\&&\mbox{-}012101201021
\end{eqnarray*}

We give solutions in the cases $n\ge 13$, $n\ne 14,15,16,20,21,22$. For each of these cases, we give an $n$-uniform square-free ternary morphism, where the images of letters are permutations of each other. This solves all but finitely many cases of the question of the author and Rampersad \cite{currie}:
Do there exist k-uniform squarefree ternary morphisms for all $k\ge 11$?
 
\section{Preliminaries}

We freely use ideas from combinatorics on words. (See \cite{lothaire} for example.) Nevertheless, we will set out some useful notions:

If word $w=ps$, we call $p$ a {\bf prefix} of $w$. 
We call $s$ a {\bf suffix} of $w$. If $p,s\ne\epsilon$, then $p$ is a {\bf proper prefix} of $w$ and $s$ is a {\bf proper suffix} of $w$.
If $w=pvs$ then $v$ is a {\bf factor} of $w$. We say that $v$ appears in $w$ with {\bf index} $|p|$. Thus, a prefix of $w$ appears in $w$ with index $0$. If $w=pvs$ where $p,s\ne\epsilon$, then we call $v$ an {\bf internal factor} of $w$.  

A {\bf square} is a word of the form $vv$, $v\ne \epsilon$. A word is {\bf square-free} if none of its factors is a square. A morphism $f$ is {\bf square-free} if $f(w)$ is square-free whenever $w$ is.
A word $xyx$ where $x$ and $y$ are letters is a {\bf length $3$ palindrome}. Thue \cite{berstelthue} studied infinite ternary square-free words with respect to the length 3 palindromes they contain. In the constructions of this paper, morphisms are proved square-free by tracking length 3 palindromes. 

Fix the ternary alphabet $\Sigma=\{0,1,2\}$.
Consider the cyclic shift morphism $\sigma:\Sigma^*\rightarrow \Sigma^*$ where $\sigma(1)=2$, $\sigma(2)=3$ and $\sigma(3)=1.$ We will say that word $z$ is a {\bf cyclic shift} of word $t$ if $z=\sigma^i(t)$ for some $i\in\{0,1,2\}$.
Call a morphism $f:\Sigma^*\rightarrow\Sigma^*$ a {\bf cyclic shift morphism} if $f(1)=\sigma(f(0))$ and $f(2)=\sigma(f(1))$. 

\begin{remark}\label{cyclic}
If $f$ is a square-free cyclic shift morphism, applying $f$ to an infinite ternary square-free word results in a ternary square-free word with a stem of length $|f(0)|$. The goal of this paper is to show that such a morphism $f$ exists with $|f(0)|=n$ for all $n\ge 13$.
\end{remark}

\section{Some square-free cyclic shift morphisms}

Consider the words
\begin{eqnarray*}
\alpha_1 &=& 21020102101201021201210212010210120102012\\
\alpha_2 &=& 2102010210120102120121020120210201210212010210120102012\\
\alpha_3 &=& 21020102101201021201210201202101201021201210212010210120102012\\
\alpha_4 &=& 210201021012010212012102012021012010210120210201210212010210120102012\\
\end{eqnarray*} The lengths of $\alpha_1,\alpha_2,\alpha_3,\alpha_4$ are respectively 41, 55, 62 and 69. Since 
\begin{eqnarray*}
41+55&\equiv&0\mbox{ (mod 4)}\\
55+62&\equiv&1\mbox{ (mod 4)}\\
41+69&\equiv&2\mbox{ (mod 4) and}\\
41+62&\equiv&3\mbox{ (mod 4),}\\
\end{eqnarray*}
\noindent we can choose distinct $q,r\in\{1,2,3,4\}$ such that $|\alpha_q|+|\alpha_r|$ is whatever we wish, modulo 4. Let distinct $q,r\in\{1,2,3,4\}$ be fixed.

\begin{remark}\label{length 3 palindromes}
Word $\alpha_q$  was obtained by computer search and has the following properties:
\begin{enumerate}
\item \label{sq-free} Word $\alpha_q$ is square-free.
\item \label{010} Prefix $\pi=210201021012010$ of $\alpha_q$ contains length 3 palindrome $010$ at indices $4$ and $12$.
\item \label{palindromes} The length 10 prefix (resp., suffix) of $\alpha_q$ contains factors $020$, $010$ and $101$; thus, each letter of $\Sigma$ is the center of some length 3 palindromic factor of the length 10 prefix (resp., suffix) of $\alpha_q.$
\item \label{02010} The only length 5 factors of $\alpha_q$ in which the same letter appears three times are $02010$ and $01020$. Each of these appears in $\alpha_q$ exactly once; the first is in the length 7 prefix, the second in the length 7 suffix.
\item \label{border} The only proper prefix (resp., suffix) of $\alpha_q$ which is a suffix (resp., prefix) of either $\alpha_q$ or of $\alpha_r$ is $2$; such a prefix/suffix must clearly begin and end with $2$; the prefix $2102$ doesn't work, and any prefix of length 7 or longer contains $02010$, and by the previous property cannot be a proper suffix.
\end{enumerate}
\end{remark}

Since $q$ was arbitrary, these properties evidently apply to $\alpha_r$ also, {\it mutatis mutandi}. Because $\alpha_q$ has suffix $\pi^R$, one checks that all these properties also apply to $\alpha_q^R$. 

\begin{remark}\label{length 7 prefix}
In any cyclic shift of $02010$ (resp. $01020$), the same letter appears 3 times; moreover, the cyclic shifts of $02010$ and $01020$ are distinct. It follows from  Remark \ref{length 3 palindromes}.\ref{02010} that if $p$ is a prefix (resp., suffix) of $\alpha_q$ or of $\alpha_r$ of length 7 or more, then no cyclic shift of $p$ is a suffix (resp., prefix) or an internal factor of $\alpha_q$ or of $\alpha_r$. 
\end{remark}

Fix $x$, $|x|\ge 2|\alpha_4|$ such that $2102012x2102012$ is a square-free word over $\Sigma$ which contains neither of $212$ and $010$ as a factor. (We will consider later for which lengths such $x$ exist.)

\begin{remark}\label{a not in x} From Remark \ref{length  3 palindromes}.\ref{palindromes}, each letter of $\Sigma$ is the center of some length 3 palindromic factor of the length 10 prefix (resp., suffix) of $\alpha_q$ (resp., $\alpha_r$). This will remain true of cyclic shifts of the length 10 prefix (resp., suffix). Since $1$ is not the center of any length 3 palindromic factor of $x$, no cyclic shift of a length 10 prefix (resp., suffix) of $\alpha_q$ (resp., $\alpha_r$) is a factor of $x$. More generally, no cyclic shift of a length 10 prefix or suffix of $\alpha_q$ or $\alpha_r$ is a factor of any cyclic shift of $x$.
\end{remark}

\begin{lemma}\label{aa}
If $i,j\in\{0,1,2\}$, $i\ne j$, word $\sigma^i(\alpha_r)\sigma^j(\alpha_q)$ is square-free.
\end{lemma}
\pf  This can be shown by a finite check.$\Box$

\begin{lemma}\label{ax}
Words $x \alpha_r$ and $\alpha_q x$ are square-free.
\end{lemma}
\pf For the sake of getting a contradiction, suppose that $zz$ is a factor of $x\alpha_r$, $z\ne \epsilon$. Since $x210201$ and $\alpha_r$ are both square-free, we can write $zz=x^{\prime\prime}210201a'$ where $x^{\prime\prime}$ is a non-empty suffix of $x$, $a'$ a non-empty prefix of $(210201)^{-1}\alpha_r$. Since $a'$ is non-empty, word $010$ is a factor of $zz$. 

\noindent {\bf Case A:} Word $010$ is a factor of $z$. Since $010$ is not a factor of $x210201$, $x^{\prime\prime}2102010$ must be a prefix of the first $z$ of $zz$, while the second $z$ of $zz$ is a factor of $a'$. This is impossible by Remark \ref{length 7 prefix}.

\noindent {\bf Case B:} Word $010$ is a not factor of $z$. It follows that either 
\begin{enumerate}
\item $z=x^{\prime\prime}21020$ or 
\item $z=x^{\prime\prime}210201$,
\end{enumerate}
 while $210201a'$ is a proper prefix of $210201021012010$.

In Case 1, the second $z$ in $zz$ is a proper prefix of $1021012010$. However, $z=x^{\prime\prime}21020$ has suffix $020$ which is not a factor of $1021012010$. This is a contradiction.

In Case 2, the second $z$ in $zz$ is a proper prefix of $021012010$. However, $z=x^{\prime\prime}210201$ has suffix $0201$ which is not a factor of $021012010$. This is a contradiction.

We conclude that $x\alpha_r$ is square-free. The same argument shows that $x^R\alpha_q^R$ is square-free, so that $\alpha_qx$ is square-free.$\Box$

\begin{lemma}\label{aqx}
Let $\beta$ be a cyclic shift of $\alpha_q$. Then $\beta$ is not an internal factor of $\alpha_q x$ or of $x\alpha_r$.
\end{lemma}
\pf Let $\alpha\in\{\alpha_q,\alpha_r\}$. Suppose for the sake of getting a contradiction that $\beta$ is an internal factor of $\alpha x$. By Lemma \ref{length 7 prefix}, $\beta$ is not an internal factor of  $\alpha$. By Remark \ref{a not in x}, $\beta$ is not a factor of $x$. Therefore, write $\beta=\alpha^{\prime\prime}x'$ where $\alpha^{\prime\prime}$ is a proper suffix of $\alpha$ and $x'$ is a proper prefix of $x$. By Remark \ref{a not in x}, $|x'|<10$. By Remark \ref{length 7 prefix}, $|\alpha^{\prime\prime}|<7$. Then $|\beta|<10+7=17$, which is absurd.

By the same arguments, $\beta^R$ is not an internal factor of $\alpha^Rx^R$, showing that $\beta$ cannot be an internal factor of $x \alpha_r.\Box$

\begin{lemma}\label{qr}
Let $\alpha$ and $\beta$ be cyclic shifts of $\alpha_q$. Let $\gamma$ and $\delta$ be cyclic shifts of $\alpha_r$. Then $\alpha$ is not an internal factor of $\gamma\beta$ and $\gamma$ is not an internal factor of $\delta\beta$.
\end{lemma}
\pf Suppose for the sake of getting a contradiction that $\alpha$ is an internal factor of $\gamma\beta$. By Lemma \ref{length 7 prefix}, $\alpha$ is not an internal factor of $\gamma$ or of $\beta$. Therefore, write $\alpha=\gamma^{\prime\prime}\beta'$ where $\gamma^{\prime\prime}$ is a proper suffix of $\gamma$ and $\beta'$ is a proper prefix of $\beta$. By Remark \ref{length 7 prefix}, $|\gamma^{\prime\prime}|,|\beta'|<7$. Then $|\alpha|<14$, which is absurd.

Similarly, $\gamma$ is not an internal factor of $\delta\beta.\Box$

Let $y=\alpha_q x \alpha_r $. Let $f:\Sigma^*\rightarrow\Sigma^*$ be the morphism given by $f(0)=y$, $f(1)=\sigma(y)$, $f(2)=\sigma^2(y)$.

\begin{theorem}\label{berstel} \cite{berstelthue} If $h$ is a uniform morphism that preserves square-free words of length 3, then $h$ is square-free.
\end{theorem}

\begin{lemma}\label{sqfree}Let $b,c,d\in\Sigma$, and let $bcd$ be square-free. Then $f(bcd)$ is square-free.
\end{lemma}
\pf Suppose for the sake of getting a contradiction that that $zz$ is a square in $f(bcd)$. We have $$f(bcd)=  \sigma^i(\alpha_q)\sigma^i(x)\sigma^i(\alpha_r)\sigma^j(\alpha_q)\sigma^j(x)\sigma^j(\alpha_r)\sigma^k(\alpha_q)\sigma^k(x)\sigma^k(\alpha_r)$$ for some $i,j,k$. By Lemmas \ref{aqx} and \ref{qr}, cyclic shifts of $\alpha_q$ only appear in $f(bcd)$ with indices $0$, $|f(b)|$ and $|f(bc)|$. If $b=d$, the cyclic shifts of $\alpha_q$ at indices $0$ and $|f(bc)|$ are identical; otherwise all three cyclic shifts of $\alpha_q$ are distinct.

\noindent {\bf Case 1. A cyclic shift of $\alpha_q$ or of $\alpha_r$ is a factor of $z$:} Suppose that a cyclic shift of $\alpha_q$ is a factor of $z$. (The case with  $\alpha_r$ is similar.) It follows that this cyclic shift of $\alpha_q$ appears twice in $zz$, hence twice in $f(bcd)$. It must therefore appear with indices $0$ and $|f(bc)|$, so that $|z|=|f(bc)|$. Then $|f(bcd)|\ge|zz|=2|f(bc)|>|f(bcd)|$. This is a contradiction.

\noindent {\bf Case 2. A cyclic shift of $\alpha_q$ or of $\alpha_r$ is a factor of $zz$:} Suppose that a factor $\beta$ of $zz$ is a cyclic shift  of $\alpha_q$. (The case with  $\alpha_r$ is similar.) By the previous case, we may suppose that $\beta$ is not a factor of $z$. We can therefore write $zz=s\beta p$ with $z=sa'=a^{\prime\prime}p$, $a'a^{\prime\prime}=\beta$. Considering the three indices where cyclic shifts of $\alpha_q$ appear in $f(bcd)$, we find that $s$ must be a suffix of a cyclic shift of $\alpha_r$. By the previous case we may assume that $|s|<|\alpha_r|$. By Lemma \ref{aqx}, $s\ne \epsilon$. Next we note that
$$|p|\le|z|=|s|+|a'|<2|\alpha_4|\le|x|.$$

Considering the indices where cyclic shifts of $\alpha_q$ appear in $f(bcd)$, we conclude that $p$ is a prefix of the cyclic shift of $x$ that follows $\beta$ in $f(bcd)$. By Lemma \ref{ab}, $p\ne \epsilon$. Recall that  $sa'=a^{\prime\prime}p$. If $|p|<|a'|-1$, then a prefix of $a^{\prime\prime}$ of length 2 or more is also a suffix of $a'$ and a  border of $\beta$. This is impossible by Remark \ref{length 3 palindromes}.\ref{border}. We conclude that $|p|\ge|a'|-1$. Similarly, $|s|\ge |a^{\prime\prime}|-1$.

If $|a'|\ge |a^{\prime\prime}|$, then $|a'|\ge|\alpha_q|/2\ge 41/2>7$. Since $|p|\ge|a'|-1$, all of $a'$, except at most a one-letter prefix, is a factor of $p$. In particular, a cyclic shift of $02010$ appears in $a'$ at index 2, and is a factor of $p$, which is a factor of $x$. This contradicts Remark \ref{length 3 palindromes}.\ref{02010}.

Similarly, if $|a'|< |a^{\prime\prime}|$, then $a^{\prime\prime}\ge|\alpha_q|/2\ge 41/2>7$. Since $|s|\ge|a^{\prime\prime}|-1$, all of $a^{\prime\prime}$, except at most a one-letter suffix, is a factor of $s$. In particular, a cyclic shift of $01020$ appears in $a^{\prime\prime}$ at index $|a_q|-2$, and is a factor of $p$, which is a factor of $x$. This contradicts Remark \ref{length 3 palindromes}.\ref{02010}.

\noindent {\bf Case 3. No cyclic shift of $\alpha_q$ or of $\alpha_r$ is a factor of $zz$:} From this case definition, together with Lemmas \ref{aqx} and \ref{ab}, we can write 
$zz= a^{\prime\prime}\chi a'$, $z= a^{\prime\prime}x'=x^{\prime\prime} a'$ where $a^{\prime\prime}$ is a suffix of a cyclic shift of $\alpha_q$, $\chi$ is a cyclic shift of $x$, $a'$ is a prefix of a cyclic shift of $\alpha_r$ and $\chi=x'x^{\prime\prime} $. Since $2102012x2102012$ is square-free, we conclude that either $|a^{\prime\prime}|\ge 7$ or $|a'|\ge 7$. Suppose $|x^{\prime\prime}|\ge |x'|$. (The other case is similar.) This implies that $|a^{\prime\prime}|\ge |a'|$, so that $|a^{\prime\prime}|\ge 7$. However,  $|x^{\prime\prime}|\ge |x'|$ implies $|x^{\prime\prime}|\ge |x|/2\ge|\alpha_q|\ge|a^{\prime\prime}|$, so that $a^{\prime\prime}$ is a factor of $x^{\prime\prime}$. We conclude that a length 7 suffix of a cyclic shift of $\alpha_q$ is a factor of $x^{\prime\prime}$, which is a factor of $\chi$. This is impossible by Remark \ref{length 7 prefix}.$\Box$
\section{Lengths for $x$}
Consider the Thue-Morse word $${\bf t}=0110100110010110\cdots$$
which is a fixed point of the morphism $h:\{0,1\}^*\rightarrow\{0,1\}^*$ given by $h(0)=01$, $h(1)=10$. We will use the following:
\begin{theorem}\label{ab}\cite{aberkane}  Let $k\ge 6$ be a positive integer. Then {\bf t} contains a factor of length $k$ of the
form $01v01$ and a factor of length $k$ of the form $01v10$.
\end{theorem}
The following has been known since Thue \cite{berstelthue}:
\begin{lemma}
Let $u$ be a factor of {\bf t} beginning and ending with 0. Write $u=01^{x_1}01^{x_2}\cdots01^{x_n}0$. Then $v=x_1x_2\cdots x_n$ is a square-free word over $\Sigma$ and neither of $010$ and $212$ is a factor of $v$. 
\end{lemma}
We will call $v$ the {\bf 1-count} of $u$.
\begin{corollary}\label{length x}
 Let $k\ge 6$ be a positive integer. There exists a square-free word $r$ over $\Sigma$ of length $4k-1$  of the form $r=2102012x2102012$ such that neither of 010 and 212 is a factor of $r$.
\end{corollary}
\pf By Lemma \ref{ab}, let $01v01$ be a factor of {\bf t} of length $k$. Then $u=h^3(01v01)=0110100110010110h^3(v)0110100110010110$. Let $r=2102012x2102012$ be the 1-count of $u$. Exactly half of the letters of $u$ are 0's, and the length of the 1-count of $u$ will be one less than the number of 0's in $u$. Thus $|r|=|u|/2-1=4k-1.\Box$ 

\section{Stems in ternary square-free words}
In our earlier definition of the cyclic shift morphism $f$, we used $\alpha_q$, $\alpha_r$ and $x$. As indicated earlier, we could choose $|\alpha_q|+|\alpha_r$ to be 96, 117, 110 or 103. Corollary \ref{length x} shows that $x$ can have any length of the form $9+4k$. This gives the following:
\begin{theorem}
There exists an $n$-uniform square-free cyclic shift morphism $f:\Sigma^*\rightarrow\Sigma^*$ for 
\begin{eqnarray*}
n&\equiv&0\mbox{(mod 4)}, n\ge 105\\
n&\equiv&1\mbox{(mod 4)}, n\ge 126\\
n&\equiv&2\mbox{(mod 4)}, n\ge 119\\
n&\equiv&3\mbox{(mod 4)}, n\ge 112
\end{eqnarray*}
such that $f(2)=\sigma(f(1))=\sigma^2(f(0)$.
In particular, such an $n$-uniform square-free morphism exists on $\Sigma$ for $n\ge 123$.
\end{theorem}
\begin{theorem}
There exists an $n$-uniform square-free cyclic shift morphism $f:\Sigma^*\rightarrow\Sigma^*$ for all $n\ge 13$, with the exception of $n=14,15,16$ and $n=20,21,22$. \end{theorem}
\pf An $n$-uniform cyclic shift morphism $f$ is specified by giving $f(0)$. Whether it is square-free is decidable by  Theorem \ref{berstel}. Exhaustive search trying all length $n$ strings as $f(0)$ disproves the existence of such morphisms for $n=14,15,16$ and $n=20,21,22$. By the previous theorem, the desired morphisms exist for $n\ge 123$. We exhibit $f(0)$ for square-free cyclic shift morphisms of length $n$, $13\le n\le 122$, $n\ne 14,15,16,20,21,22$ in the Appendix. 
Applying such an $f$ to any infinite square-free word over $\Sigma$ 
gives an infinite square-free ternary word with an $n$-stem factorization.
\begin{theorem}\label{stem}
An infinite square-free ternary word with an $n$-stem factorization exists for any $n\ge 23$.
\end{theorem}
 
\section{Appendix: Square-free cyclic shift morphisms of length $n$}
\begin{tabular}{|r|c|}
\hline
$n$&$f(0)$\\\hline
13&2101201021012\\

17&21020102120102012\\
18&210201021202102012\\
19&2102010210120102012\\

23&21012010201210201021012\\
24&210120102012021201021012\\
25&2101201020120210201021012\\
26&21012010201202101201021012\\
27&210120102012102010212021012\\
28&2102010212012101202120102012\\
29&21020102101210201202120102012\\
30&210201021012021201210120102012\\
31&2102010210120102120210120102012\\
32&21020102101201021201210120102012\\
33&210201021012010212012102120102012\\
34&2102010210120102012101202120102012\\
35&21020102101201020121020102120102012\\
36&210201021012010212012101202120102012\\
37&2102010210120102012021020102120102012\\
38&21020102101201020120210201210120102012\\
39&210201021012010212012102120210120102012\\
40&2102010210120102120121020120210120102012\\
41&21020102101201020120210121020102120102012\\
42&210201021012010201202101210201202120102012\\
43&2102010210120102012021020102101202120102012\\
44&21020102101201020120210120102101202120102012\\
45&210201021012010201202101201021012102120102012\\
46&2102010210120102012021012010212012102120102012\\
47&21020102101201020120210121020102101202120102012\\
48&210201021012010201202101201021202101202120102012\\
49&2102010210120102012021012010210121020102120102012\\
50&21020102101201020120210120102101210201202120102012\\
51&210201021012010201202101201021012021020102120102012\\\hline
\end{tabular}

\begin{tabular}{|r|l|}
\hline
$n$&$\hspace{2in} f(0)$\\\hline
52&2102010210120102012021012010210120210201202120102012\\
53&21020102101201020120210120102101202102010210120102012\\
54&210201021012010201202101201021012021020120210120102012\\
55&2102010210120102012021012010210121020102101202120102012\\
56&21020102101201020120210120102101202102012101202120102012\\
57&210201021012010201202101201021012021020102101202120102012\\
58&2102010210120102012021012010210120210201021012102120102012\\
59&21020102101201020120210120102101202102010212012102120102012\\
60&210201021012010201202101201021012021020120212010210120102012\\
61&210201021012010201202101201021012021020102120210120212010201\\&-2\\
62&210201021012010201202101201021012021020102101210201021201020\\&-12\\
63&210201021012010201202101201021012021020102101210201202120102\\&-012\\
64&210201021012010201202101201021012021020102101202120121012010\\&-2012\\
65&210201021012010201202101201021012021020102101201021202101201\\&-02012\\
66&210201021012010201202101201021012021020102101201021201210120\\&-102012\\
67&210201021012010201202101201021012021020102101201021201210212\\&-0102012\\
68
&210201021012010201202101201021012021020102101201020121012021\\&-20102012\\
69
&210201021012010201202101201021012021020102101201020121020102\\&-120102012\\
70
&210201021012010201202101201021012021020102101201021201210120\\&-2120102012\\
71&210201021012010201202101201021012021020102101201020120210201\\&-02120102012
\\
72
&210201021012010201202101201021012021020102101201020120210201\\&-210120102012\\
73
&210201021012010201202101201021012021020102101201021201210201\\&-02101201020
12\\
74
&210201021012010201202101201021012021020102101201021201210201\\&-20210120102
012\\\hline
\end{tabular}
\begin{tabular}{|r|l|}
\hline
$n$&$\hspace{ 2 in}f(0)$\\\hline
75
&210201021012010201202101201021012021020102101201020120210121\\&-02010212010
2012\\
76
&210201021012010201202101201021012021020102101201020120210121\\&-02012021201
02012\\
77
&210201021012010201202101201021012021020102101201020120210201\\&-02101202120
102012\\
78
&210201021012010201202101201021012021020102101201020120210201\\&-02101210212
0102012\\

79
&210201021012010212012102012021012010201202101210201202120121\\&-02120102101
20102012\\
80
&210201021012010212012102012021012010201202101210201021202101\\&-20212010210
120102012\\81
&210201021012010212012102012021020121021201021012102010212021\\&-01202120121
0120102012\\
82
&210201021012010212012102012021020121021201021012021201021202\\&-10120212012
10120102012\\
83&2102010210120102120121020120210201210212010210120210201021\\&-2021012021201
210120102012\\
84&2102010210120102120121020120210201210212010210120102012102\\&-0102101202120
1210120102012\\
85&2102010210120102120121020120210201210212010210120102012021\\&-2010210120212
01210120102012\\
86&2102010210120102120121020120210201210212010210120102012021\\&-0201021012021
201210120102012\\
87&2102010210120102120121020120210201210212010210120102012021\\&-0120102101202
1201210120102012\\
88&2102010210120102120121020120210201210212010210120102012102\\&-0102120210120
21201210120102012\\
89&2102010210120102120121020120210201210212010210120102012021\\&-2010212021012
021201210120102012\\
90&2102010210120102120121020120210201210212010210120102012021\\&-0121020102101
2021201210120102012\\
91&2102010210120102120121020120210201210212010210120102012021\\&-0120102120210
12021201210120102012\\
92&2102010210120102120121020120210201210212010210120102012021\\&-0121021201021
012021201210120102012\\\hline
\end{tabular}

\begin{tabular}{|r|l|}
\hline
$n$&$\hspace{ 2in}f(0)$\\\hline
93&2102010210120102120121020120210201210212010210120102012021\\&-0201210120102
1012021201210120102012\\
94&2102010210120102120121020120210201210212010210120102012021\\&-0121020102120
21012021201210120102012\\
95&2102010210120102120121020120210201210212010210120102012021\\&-2010201210201
021012021201210120102012\\
96&2102010210120102120121020120210201210212010210120102012021\\&-2010201210120
1021012021201210120102012\\
97&2102010210120102120121020120210201210212010210120102012021\\&-0121020120212
01021012021201210120102012\\
98&2102010210120102120121020120210201210212010210120102012021\\&-0120102101210
201021012021201210120102012\\
99&2102010210120102120121020120210201210212010210120102012021\\&-0120102120121
0201021012021201210120102012\\
100&21020102101201021201210120102012021012010201210120102101\\&-20210201021012010201210201021201210120102012\\
101&21020102101201021201210120102012021012010201210120102101\\&-202102010210120102012101201021201210120102012\\
102&21020102101201021201210120102012021012010201210120102101\\&-2021020102101201020120210201021201210120102012\\
103&21020102101201021201210120102012021012010201210120102101\\&-20210201021012010201202101201021201210120102012\\
104&21020102101201021201210120102012021012010201210120102101\\&-202102010210120102012021020121021201210120102012\\
105&21020102101201021201210120102012021012010201210120102101\\&-2021020102101201021202101210201021201210120102012\\
106&21020102101201021201210120102012021012010201210120102101\\&-20210201021012010201202101210201021201210120102012\\
107&21020102101201021201210120102012021012010201210120102101\\&-202102010210120102012021012102012021201210120102012\\
108&21020102101201021201210120102012021012010201210120102101\\&-2021020102101201020120210201021012021201210120102012\\
109&21020102101201021201210120102012021012010201210120102101\\&-20210201021012010201202101201021012021201210120102012\\
110&21020102101201021201210120102012021012010201210120102101\\&-202102010210120102012021012010210121021201210120102012\\\hline
\end{tabular}

\begin{tabular}{|r|l|}
\hline
$n$&$\hspace{ 2in}f(0)$\\\hline
111&21020102101201021201210120102012021012010201210120102101\\&-2021020102101201020120210121020121012021201210120102012\\
112&21020102101201021201210120102012021012010201210120102101\\&-2021020102101201020120210121020102101202120121012010201\\&-2\\
113&21020102101201021201210120102012021012010201210120102101\\&-2021020102101201020120210120102120210120212012101201020\\&-12\\
114&21020102101201021201210120102012021012010201210120102101\\&-2021020102101201020120210120102101210201021201210120102\\&-012\\
115&21020102101201021201210120102012021012010201210120102101\\&-2021020102101201020120210120102101210201202120121012010\\&-2012\\
116&21020102101201021201210120102012021012010201210120102101\\&-2021020102101201020120210120102101202102010212012101201\\&-02012\\
117&21020102101201021201210120102012021012010201210120102101\\&-2021020102101201020120210120102101202102012021201210120\\&-102012\\
118&21020102101201021201210120102012021012010201210120102101\\&-2021020102101201020120210120102101202102012102120121012\\&-0102012\\
119&21020102101201021201210120102012021012010201210120102101\\&-2021020102101201020120210120102101210201210120212012101\\&-20102012\\
120&21020102101201021201210120102012021012010201210120102101\\&-2021020102101201020120210120102101202120102012021201210\\&-120102012\\
121&21020102101201021201210120102012021012010201210120102101\\&-2021020102101201020120210120102101202102012101202120121\\&-0120102012\\
122&21020102101201021201210120102012021012010201210120102101\\&-2021020102101201020120210120102101202102010210120212012\\&-10120102012\\\hline
\end{tabular}

\end{document}